\documentclass{article}
\usepackage[utf8]{inputenc}
\usepackage[T1]{fontenc}
\usepackage{amsmath}
\usepackage{amsfonts}
\usepackage{latexsym}
\usepackage{amssymb}
\usepackage{times}
\usepackage{ifpdf}
\usepackage{makeidx}
\usepackage{graphicx}
\ifpdf
 \usepackage[pdftex,colorlinks]{hyperref}
\else
 \usepackage[ps2pdf,breaklinks=true,colorlinks=true,linkcolor=red,citecolor=green]{hyperref}
 \fi

\newcommand{\Z}{{\mathbb{Z}}}

\newcommand{\Q}{{\mathbb{Q}}}

\newtheorem{thm}{Theorem}
\newtheorem{prop}[thm]{Proposition}

\title{Certifying a probabilistic parallel modular algorithm for rational univariate representation.}

\author{Bernard Parisse\\Institut Fourier\\UMR 5582 du
  CNRS\\Universit\'e de Grenoble Alpes}
\date{September 2021}

\begin{document}

\maketitle

\begin{abstract}
This paper is about solving polynomial systems. 
It first recalls how to do that efficiently
with a very high probability of correctness by
reconstructing a rational univariate representation (rur)
using Groebner revlex computation, Berlekamp-Massey algorithm and Hankel 
linear system solving modulo several primes in parallel. Then it
introduces a new method (theorem \ref{prop:check})
for rur certification 
that is effective for most polynomial systems.

These algorithms are implemented in
\href{https://www-fourier.univ-grenoble-alpes.fr/~parisse/giac.html}{Giac/Xcas} (\cite{giac})
since version 1.7.0-13 or 1.7.0-17 for certification, it has (July 2021)
leading performances on multiple CPU, at least for an
open-source software. 
\end{abstract}

\section{Introduction}
Polynomial system solving can be performed by doing several
eliminations. If the variables are $x_1,..,x_n$, after eliminating
$x_1,...,x_{n-1}$, one has
to solve one (large degree) univariate polynomial, and then one
finds other unknowns by back substitution, gcd computations and univariate polynomial
solving. Unfortunately this method (a regular chain type method), requires
building tower of algebraic extensions over $\Q$, which is computation
intensive.

It is more efficient to build one algebraic extension of $\Q$ (or more
if the system factors)  such
that all components of the solutions of the system will live in this
extension. This can be performed by computing a Gr\"obner basis of
$I$, the ideal spanned by the multivariate polynomials of the system. 
Then, if the ideal is 0 dimensional, select one variable (say $x_n$), find the minimal
polynomial of this variable. If this polynomial $m$ has the right
degree (the dimension of the polynomials modulo the ideal as a vector space)
and is square free,  for all other variables, find $P_i$ such that
$x_i-P_i(x_n)$ is inside the ideal. Then the system solutions are
$P_1(x_n),P_2(x_n),...,x_n$ for all roots $x_n$ of $m$.

Rouillier 
(\cite{rouillier1999solving})
found that it is more efficient to compute $Q_i$ such
that $x_i-Q_i(x_n)/m'(x_n)$ is inside the ideal\footnote{which means that
  $m' x_i -Q_i$ is in the ideal. Since $m$ and $m'$ are coprime and
  $m$ is in the ideal, a
polynomial $P$ belongs to the ideal if and only if $m'P$ is in the ideal.}, and this is called
rational univariate representation.

If for all unknowns, the minimal polynomial degree is too small, a
linear separating form (linear combination of the $x_i$), must be
found such that the minimal polynomial is of degree the dimension of
the vector space of the polynomials modulo the ideal. If the
ideal is not radical 
(i.e. if there is a polynomial $P$ that does not belongs to $I$ but
$P^k \in I$ for some $k>0$), the method must be adapted.

Modular algorithms are well known techniques in CAS to make efficient
computations, they are also good candidates for parallelization. In
our context, algorithms were presented early, like in \cite{noro1999modular}
(\href{https://www-polsys.lip6.fr/~jcf/Papers/master2.pdf}{A Modular
  Method to Compute the Rational Univariate Representation of
  Zero-dimensional Ideals}). For more recent results, see
\cite{berthomieu2021msolve}.

We  will review some of these algorithms as well as some algorithms of Faug\`ere
and Mou
(\cite{faugere2017sparse}) for rur computation in $\Z/p\Z$ in Section
2 (to be more precise, we will compute 
the rur for the radical ideal spanned by the polynomials of the
system). In subsection \ref{sec:certif} we will precise how we can
certify a rur.
Section 3 will give more informations on the implementation inside
Giac/Xcas and
gives some benchmarks.

\section{Algorithms for RUR computations}

\subsection{Gr\"obner basis over $\Z/p\Z$ (degree rev. lex. ordering)}
For a basis computation without additional hypothesis, it seems that
F4 (\cite{F99a}) is a very good algorithm choice. Since the same basis is computed several
times for different primes, we can store some informations during the
first run, like s-pairs reducing to 0, in
order to speed up computation for the next primes, 
this is described in more details in \cite{parisse2013probabilistic}
(\href{https://arxiv.org/abs/1309.4044}{A probabilistic and
  deterministic modular algorithm for computing Groebner basis over
  $\Q$}).

\subsection{From Gr\"obner basis to RUR over $\Z/p\Z$}
If the ideal $I$ spanned by the polynomials of the system
 is 0-dimensional, the polynomials modulo $I$ belong to $V$, a vector space of
finite dimension $d$. We can compute a basis of $V$ by collecting all
the monomials smaller than the leading ones in $G$.

The reduction of $1,x_n,..,x_n^d$ with
respect to the Gr\"obner basis is not free, there is a minimal polynomial $m$
of degree at most $d$ such that $m(x_n)=0 \pmod I$. Computing $m$ is
a linear algebra kernel computation (for a matrix with columns the
components of the reduction mod $I$ of $x_n^i$). This is an $O(d^3)$
computation with a naive Gauss pivoting method. Fortunately, it can be
computed faster, by observing that if
$$ \sum_i m_i x_n^i = 0 \pmod I $$
then it's still true after multiplication by a power of $x_n$, and
therefore the scalar product of a fixed vector and the reduced vector is 0, then the
$m_i$ coefficients can be found by mean of the Berlekam-Massey algorithm,
using the half-gcd fast version in $O(d\log(d))$ operations
(see
e.g. \cite{yap2000fundamental},
\href{http://webéducation.com/wp-content/uploads/2018/11/Fundamental-Problems-in-Algorithmic-Algebra.pdf}
{Yap}), 
once the scalar products are computed.

There is still a naive $O(d^3)$ part in this algorithm, computing the
$x_n^k \pmod I$. This is done by computing the matrix of the
multiplication by $x_n \pmod I$ in the basis of $V$, by reducing all
monomials of this basis multiplied by $x_n$. In many situations,
the multiplication by $x_n$ will give a monomial of the basis of $V$,
and the corresponding column of $M$ is trivial (one 1 and all other coefficients
0). Or the multiplication by $x_n$ will return the leading monomial of one element of $G$ and
the reduction is trivial (take the opposite of all remaining monomials
of this element of $G$). The remaining products by $x_n$ must be reduced mod $G$,
this can be done simultaneously for all these products like in the F4
algorithm. The multiplication of a vector by $M$ is still $O(d^2)$,
but it becomes faster if the matrix has many trivial columns, and it's
a simple operation that can benefit from the CPU instruction set.

In a generic situation, the minimal polynomial of $x_n$ 
is of maximal degree $d$ the dimension
of $V$ the vector space of polynomials modulo $I$ and every element
of $V$ (i.e. any polynomial modulo the ideal $I$) can be expressed as a polynomial in $x_n$, of
degree $<d$. Finding the polynomial
corresponding to $x_1,..,x_{n-1}$ will give the solution to the initial
system. It can be seen as a linear system of matrix with columns the
powers 0 to $d-1$ of $x_n$ reduced modulo $I$ and with second member
$x_i$ modulo $I$. Which means solving $n-1$ systems
with the same $d \times d$ matrix, where $n<<d$, at a $O(d^3)$ cost
(naive algorithm). Fortunately, this can be improved, by observing
that if $x_i=P_i(x_n) \pmod I$ then 
$x_n^k x_i=x_n^k P_i(x_n) \pmod I$, then one can do the
scalar product of this equation with any fixed vector, and
get a linear equation in the coefficients of $P_i$ that we are
computing. Doing that for $k=0,...,d-1$ will bring a linear system
with a structured matrix. This matrix is named a Hankel matrix, it's coefficients
are already computed: it's the coefficients of the Berlekamp-Massey
algorithm that returned the minimal
polynomial of $x_n$. Hankel matrices can be inverted using an extended
GCD and a Bezoutian matrix,
cf. for example
\href{https://en.wikipedia.org/wiki/Hankel_matrix}{wikipedia}
and the cost for computing a Bezoutian matrix is $O(d^2)$, cf. e.g.
\cite{chionh2002fast} 
(\href{https://citeseerx.ist.psu.edu/viewdoc/download?doi=10.1.1.85.3710&rep=rep1&type=pdf}{Fast
  computation of the Bezout and Dixon resultant matrices})

\subsection{Non generic situations}
If the minimal polynomial of $x_n$ is not of maximal degree, one can
try the other monomials $x_1$ to $x_{n-1}$ and if it does not work, a 
random linear combination of $x_1,...,x_n$, this is called a
separating linear form. Finding a linear separating form may be hard,
see. for example \cite{bouzidi2016solving} for bivariate systems.

If the minimal polynomial is of maximal degree $d$ but is not squarefree, 
then the ideal $I$ is not
radical, in that case one can add the squarefree part of the minimal
polynomial
to the generators of the ideal and compute a new Gr\"obner basis,
until the ideal is radical (it is not mandatory to reconstruct a
radical rur, but it is convenient for a software like Giac/Xcas where
we are interested in system solutions and not multiplicities). Cf. 
\href{https://hal.inria.fr/hal-00807540}{Faug\`ere and Mou}
(\cite{faugere2017sparse})
for alternative methods.

\subsection{Rational reconstruction}
It is of course possible to compute a Gr\"obner basis over $\Q$ and
run the same kind of computations, but field operations in $\Q$
are not performed in $O(1)$ time, that's why a multi-modular algorithm
is most of the time more efficient.

 The first step is to cancel
denominators so that the coefficients belong to $\Z$. 
A prime $p$ is said to be a good
reduction prime if the steps of the computation over $\Z/p\Z$ 
are the reduction modulo $p$ of the steps over $\Q$. This is true
if the leading monomials of the s-pairs do not cancel mod $p$, 
if the basis of the vector space of the polynomials modulo the ideal remains the
same, if the degree of the 
minimal polynomial of the separating linear form remains $d$.
Therefore before trying to reconstruct a rur in $\Q$ from several
primes, we must check the consistency of these primes. Two primes
are compatible if the leading monomials of the Gr\"obner basis have
the same power exponents. If one prime has
a gbasis with less elements, or a basis of $V$ with less elements,
it must be discarded.

My estimate for the probability to have a bad prime for a today-large
computation is less than \verb|1e-4| (with about \verb|1e4| leading
coefficients for a prime of size about \verb|5e8| in Giac), hence if a few thousands
primes are required to stabilize the computation over $\Q$, the
probability to  meet one bad prime would be less than 0.1.

Reconstruction is done coefficient by coefficient using Farey algorithm. If the reconstructed
rur modulo the next prime $p$ matches the computation over 
$\Z/P\Z$ (where $P$ is the product of the previous primes), 
then the probability of a bad rur reconstruction is very low and can be as
low as desired by checking with a few other primes rur computations. Certifying
a rur is somewhat costly, because we must either reduce
the rur elements modulo the $\Q$-gbasis (which must be certified as
well) or check that the solutions verify the initial system. We will
show in the next section that the second method can be done
efficiently by reformulating a large computation with univariate fractions with
rational coefficients as a large computation with integer coefficients.

\subsection{Certifying a rur} \label{sec:certif}
In this section, we explain how we can certify that
our rur gives all the solutions of the original system if the ideal is
radical.
 
The idea is very simple: 
just replace in the original system 
all variables by their rur fraction representation
$Q_i(x_n)/m'(x_n)$ and check if we get 0. Then we are sure
that all roots we will compute with the rur are solutions of the
system. In the other direction, we must check that
we do not miss solutions. If the ideal is radical, this is a consequence of
theorem 7.1 of Arnold \cite{Arnold2003403} that states that 
in order to check the reconstruction of a gbasis by a modular
algorithm, we must check that the reconstruction is a Groebner basis (i.e.
check that all s-polynomials of a pair reduces to 0)
and that the initial generators belong to the ideal spanned by the
reconstruction (that's precisely what we do in the substitution
check).

Here we just have to translate rur properties into Groebner basis
properties. We add a variable $t$ 
that is the common separating linear form common to all primes used
for  reconstruction and add $t$
minus the linear form to the ideal generators (generically
$t=x_n$ and we add $t-x_n$ to the initial polynomial system).
Then the set $S$ of $(P_i)_{1\leq i \leq n}:=x_i- \mbox{rem}(Q_i(t) (m'(t) [m])^{-1} ,m)$ and $m(t)$
is a Groebner basis with respect to lexicographic ordering
$x_1,..,x_n,t$ modulo each prime used for reconstruction, and
it is also a Groebner basis over $\Q$, 
indeed if we compute the s-polynomial of two elements of $S$, we get
if $m$ is not in the pair~:
\begin{eqnarray*}
x_jP_i-x_iP_j
&=&  x_j(x_i-Q_i(t) (m'(t) [m] )^{-1} [m])-
x_i((x_j-Q_j(t) (m'(t) [m] )^{-1} [m])\\
&=&x_i Q_j(t) (m'(t) [m] )^{-1} [m]
- x_j Q_i(t) (m'(t) [m] )^{-1} [m]\\
&\rightarrow& 0
\end{eqnarray*}
or if $m(t)$ is one of the two polynomials, then 
\begin{eqnarray*}
t^d P_i - x_i m
&=& (t^d -m)x_i- t^d Q_i(t) (m'(t) [m])^{-1} [m]) \\
&\rightarrow & (t^d-m) (Q_i(t) (m'(t) [m])^{-1} [m] ) - 
t^d  Q_i(t) (m'(t) [m])^{-1} [m]\\
&\rightarrow& 0
\end{eqnarray*}

In order to avoid rational computations, we write $Q_i$ as a quotient
$\tilde{Q}_i/q_i$ where $\tilde{Q}_i \in \Z[X]$ and $q_i \in \Z$ and
$m'=\tilde{D}/\tilde{d}$. The size of the coefficients is proportionnal to
$N$ the number of primes that were necessary to reconstruct the rur over
$\Q$. The degree is $\leq d$.

Let $P_j \in \Z[x_1,...,x_n]$ be a polynomial equation in the system, of total degree
$\delta$. We can perform all
computations in $\Z[X]$ by multiplying the equation by $m'^\delta$. 

Indeed, replacing values in a monomial $c_k \prod_l x_l^{\alpha_l}$ will lead to
computing
$$ c_k (\tilde{D}/\tilde{d})^{\delta-\sum \alpha_l} \prod_l
(\tilde{Q}_l/q_l)^{\alpha_l}
= \frac{c_k \tilde{D}^{\delta-\sum \alpha_l} \prod_l \tilde{Q}_l^{\alpha_l}
}{\tilde{d}^{\delta-\sum \alpha_l} \prod_l q_l^{\alpha_l}} $$
We will have less than $\delta$ products of a polynomial of degree less
than $\delta d$ with a polynomial of degree less than $d$ and
coefficients size are bounded by $\delta N$, this is
$O(\delta^3 dN)$ (up to logarithmic terms)
using FFT. 

With a divide and conquer
product algorithm, the cost becomes $O(\delta^2 d N) $.
Indeed if $T(\delta)$ is a bound for this cost, for $\delta$ even
$$ T(\delta) \leq 2 T(\frac{\delta}{2}) + M(\frac{\delta}{2},d,N)$$
where $M(\delta,d,N)$ is the cost to multiply to univariate polynomials
of degree $\delta d$ with integer coefficients of size $\leq \delta N$.
\begin{prop}
Let $M(\delta,d,N)$ be the cost to multiply two univariate polynomials
of degree $\delta d$ with integer coefficients of size $\leq \delta N$.
Then $M(\delta,d,N)\leq C\delta^2 d N$ where logarithmic terms are inside
$C$.
\end{prop}
Proof~: if the coefficients are smaller than $B$ then the product
coefficients are smaller than $\tilde{B}=\delta d B^2$. The product degree
is $\leq 2\delta d$.
Then we make $O(\log(\tilde{B}))$ FFT product of degree $\leq 2\delta d$ 
modulo small primes
and recover the integer polynomial product by chinese remaindering.

If $N\geq d$\footnote{In the
benchmarks section below, the value of $N$ should be multiplied by 
the bitsize of primes, i.e. 29},
we can choose $l$ and the smallest possible $r$ such that 
$$ 2\delta d \geq 2^l>\delta d , \quad  2^{r2^l} > \tilde{B} $$
and make a unique FFT product of the polynomials
modulo $n=2^{r2^l}+1$ (so that
reduction modulo $n$ is easy), using $2^r$ as a 
$2^{l+1}$ root of unity. The ring operations are done in
$O(\log(n))=O(\log(\tilde{B})=2dN)$ operations 
and there are up to logarithmic terms $O(2^{l+1})$ operations.

Hence for $\delta$ a power of 2~:
\begin{eqnarray*}
 T(\delta) &\leq &2 T(\frac{\delta}{2}) + C d\delta^2 N \\
&\leq & 4T(\frac{\delta}{4})+Cd\frac{\delta^2}{2}N+C d \delta^2 N  \\
&\leq & ...\\
&\leq & CdN \delta^2 ( 1 + \frac{1}{2}+ \frac{1}{4} +...) 
\end{eqnarray*}

Then we add monomials by applying~:
$$ \frac{A}{a} +\frac{B}{b}= \frac{A \frac{b}{g}  + B
  \frac{a}{g}}{g\frac{a}{g}\frac{b}{g}} , \quad
g=\mbox{gcd}_{\Z}(a,b)$$
For the cost analysis, observe that if $q$ is the lcm of the denominators of
the $Q_j$ polymomials, then $a$ and $b$ are divisors of $(q\tilde{d})^\delta$, we
could therefore replace monomial additions above by monomial
additions over $\Z[X]$. The coefficients of $A$ and $B$ would be
multiplied by at most $(q\tilde{d})^\delta$, this adds $\delta N$ to a size already
$O(\delta N)$, and the size remains an $O(\delta N)$. Hence a monomial
addition cost is in $O(\delta N \delta d)$.

The total cost of computing $P_j$ is therefore an $O(\delta^2 d N l(P_j))$ where
$l(P_j)$ is the number of monomials of $P_j$ (assumed to
be represented as a sparse distributed polynomial), $\delta$ the total
degree of $P_j$, $d$ the dimension of the vector space $V$ (the
polynomials modulo the ideal),
$N$ the number of primes\footnote{A more precise estimate is $O(dN\sum_{\mbox{monomials}}
\mbox{total degree(monomial)}^2  )$}.

The memory required is proportionnal to $O(\delta^2 d N )$.
If $\delta$ is large, the bottleneck for checking will be
memory instead of time, since it will become (much) more than
the memory required to store the rur in $O((n+2) d N )$ where $n$ is
the number of variables, especially if this step is parallelized~:
care must be taken to bound the number of parallel threads running
simultaneously ($\delta^2 \cdot \mbox{\#threads}$ should be 
of the same size order than $n$).
Another option (not tested) would be to adopt
a dense recursive representation for the polynomials of the system if
it is dense.

And at the end we compute the euclidean division with the primitive
part of the minimal polynomial $m$. Since it is highly probable that
the remainder of the division is 0, the quotient should belong to $\Z[X]$, therefore
 we can reconstruct the quotient in $\Z[X]$ by a multi-modular
 algorithm (with fast modular univariate division algorithms for each
 prime) and we do the final check by a product. Therefore if
the certification does not fail, this
division has the same cost as multiplying two polynomials of
degree $\delta d$ and $d$ and coefficient sizes $\delta N$ and $N$,
again an $O(\delta^2 d N)$ up to logarithmic terms
and the check can be performed efficiently. We get~:
\begin{thm} \label{prop:check}
The time cost of a successfull check by
substitution that the rur (rational univariate
representation) is a solution of a polynomial system
may be bounded by $O(\delta^2 d N l)$ up to logarithmic
terms, where $\delta$ is the total degree of the polynomial system,
$d$ is the dimension of $V$ (the vector space of polynomials modulo
the ideal $I$ spanned by the polynomials of the system), $N$ the
number of (fixed bit size) primes required for modular reconstruction of the rur, and $l$
the number of monomials in the polynomial system. The memory cost
may be bounded by $O(\delta^2 d N)$.

If the substitution check is successfull for all the polynomials of
the system and if the ideal is radical, then the rur is certified (if
the ideal is not radical, solutions are certified, but it is not
proved that additional solutions do not exist).
\end{thm}
If the ideal is not radical, the probability to miss solutions is
extremly small, because it would imply that 
\begin{itemize}
\item either all primes used for
reconstruction are bad primes and have at least one common leading
coefficient(s) of the s-polynomials used to compute the gbasis is 0 modulo these primes.
\item or if $m$ is the minimal polynomial of the separating linear
  form and $P=m/gcd_{\mathbb{Q}}(m,m')$ it's square-free part,
then resultant$(P,P')$ is
0 modulo all these primes.
\end{itemize}
 For example in the benchmarks section below, \verb|phuoc| is the only
 example that is not radical.
Reconstruction requires 781 primes larger than 5e8, the probability
to miss solutions is smaller than \verb|1e-6700|.

\section{Giac/Xcas implementation and benchmarks}

\subsection{Implementation}
\begin{itemize}
\item
Step 1: compute the gbasis for revlex order modulo a prime $p$ .
Giac implementation details of the gbasis algorithm with learning are described in
\cite{parisse2013probabilistic}. If $p$ is not the first prime,
compare if the current prime is compatible with previous one (leading
monomials of the gbasis must be the same), if not discard it (or all
previous primes).\\
Following a suggestion of F. Rouillier, step 1 can be replaced by a modular reduction
of the gbasis over $\Q$ if it has been already computed. If
reconstructing the gbasis over $\Q$ requires
less primes than reconstructing the rur, this speeds up a little
bit the computation (up to a factor 2 on Katsura examples below) but
it requires more memory. 
However the rur reconstruction requires often much less primes 
than the gbasis reconstruction (it probably means that representing variables
as fractions instead of polynomials is really effective in terms
of coefficient sizes, in other words that the rur is really efficient)
and doing that would require much more time. \\
 Giac/Xcas has a fine-tuning command \verb|rur_gbasis(n)|
for that purpose, if \verb|n==0|, no reconstruction of the gbasis (default), if
\verb|n==1|, reconstruction of the gbasis but leave as soon as
the rur is reconstructed, 
and if \verb|n>1|, reconstruction happens only if the number of monomials of the gbasis
is less than \verb|n|.
\item
Step 2 (for the first prime): find the dimension and a basis of $V$ made of monomials.
We collect the leading monomials of the gbasis. For every variable
$x_i$ we search a leading monomial $x_i^{d_i}$ that is a power of this variable. This will
bound any monomial exponent in $V$ by $(d_1,...,d_n)$. The dimension
$d$ of $V$ is
smaller than $D$ the product of $d_i$. For any integer $0\leq i<D$,
write $i$ in multi-basis $d_1,...,d_n$
$$ i=(..(i_1d_2+i_2)d_3+...+i_{n-1})d_{n}+i_n, \quad 0\leq i_k<d_k$$
and check if $x_1^{i_1}...x_n^{i_n}$ is greater than a leading
monomial of the gbasis, if not add it to the basis.

\item
Step 3: compute the matrix of multiplication by $x_n$ in our basis
of $V$. If $x_n$ times the monomial is itself a monomial in $V$, we do
not store a column with one 1 and $d-1$ zeros, instead we store the
pairs of indices of the monomials, this is a mixed storage (dense
part/sparse part).

\item
Step 4: compute the coefficients of the Hankel matrix (the dense
multiplication part can take advantage of the AVX2 instruction set if
available). For the dense part of the multiplication,
we avoid divisions (except the final one) by computing representants
$0\leq r <p^2$, after an addition of a multiplication of coefficients
in $[0,p)$ we substract $-p^2$ and add $p^2$ if the result is negative
without testing (for a 63 bits signed integer $i$ this is done by 
\verb|i += (i>>63) & p2|).
In Giac, we do that for 4 additions at a time (using representants in
$[0,4p^2)$ where $p<2^{29}$).

\item
Step 5: find $m$ the minimal polynomial of $x_n$ by the halfgcd
Berlekamp-Massey algorithm. If it is not of
maximal degree $d$, replace $x_n$ by another variable $x_1,..,x_{n-1}$
and go to step 3. If none of the variables fill the degree condition,
try with a random integer linear combination of $x_i$. The separating
linear form will be recorded for further primes.

\item
Step 6: if the minimal polynomial $m$ is not squarefree, add the square
free part \verb|m/gcd(m,m')| to the gbasis, and go to step 1.

\item
Step 7: find the polynomials $P_i$ such that $x_i-P_i(x_n)=0 \pmod I$
by solving Hankel systems (using fast inversion of the Hankel matrix
with bezoutians). 

\item
Step 8: compute $Q_i=P_i m' \pmod m$

\item
Step 9: (if not at the first prime)
Check if the Farey rational reconstruction for previous primes matches
this prime for $Q_i \pmod p$ (check for a few monomials before
doing a complete reconstruction check). If so, return the Farey reconstruction.
Otherwise, apply the Chinese Remainder Theorem for $Q_i \pmod p$  and
previous primes and go to step 1 for a next prime.

\item Certification. The default is to certify all equations.
 Running \verb|rur_certify(0)|will not run any certification,
\verb|rur_certify(1)|will run all certifications
while \verb|rur_certify(n)| will certify only equations of total degree
 $\delta <n$ (for $n>1$). 
For example running \verb|rur_certify(19)| for the
\verb|phuoc| example below will only certify one of the 22 equations
of the system (requires about one day of CPU) since the total degrees
are 22 (1 occurence), 20 (1 occurence), 19 (19 occurences) and 18 (1
occurence, with 1330 monomials).
\end{itemize}

Steps 1 to 8 can be parallelized. Trying to parallelize
step 9 does not speed up the computation
because it requires a lot of memory allocations, and this seems
to always be thread-mutually exclusive.

Certification can be parallelized but is limited to 6 threads by
default to spare memory. The maximal number of threads $t$ for this step
is configurable by running the command \verb|rur_certify(-t)|.

\subsection{Benchmarks}
\begin{itemize}
\item Giac/Xcas 1.7.0-17
timings are for a rur computation with 
AVX2 enabled, on an Intel(R)
Xeon(R) CPU E5-2640 v3 @ 2.60GHz.
The computation were run with 16 threads in
parallel, with a few exceptions with 8 threads in order to spare memory.
Certification is run with 6 threads in order to spare memory.
\item In order to compile 
\href{http://www-fourier.univ-grenoble-alpes.fr/~parisse/giac_compile.html}{Giac/Xcas} 
with AVX2 support with gcc,
install \href{https://github.com/vectorclass}{VCL vectorclass} by Agner Fog  and run\\
  \verb|export CXXFLAGS='-O2 -g -mfma -mavx2 -fabi-version=0'|\\
before running \verb|./configure| in the Giac/Xcas source root directory.

\item The Giac/Xcas script files for these benchmarks are available
\href{https://www-fourier.univ-grenoble-alpes.fr/~parisse/giac/rur_examples.tgz}{here}
\item The threads column is the number of threads for this
  computation.
\item The next two columsn are real time for the computation without
  certification and for certification. For * examples, gbasis reconstruction
was enabled\footnote{For a benchmark family, testing both methods for
  small benchmarks is a good hint on what should be done for large ones} 
with the \verb|gbasis_rur(1)| command, for other examples, it was not.
\item msolve timings are for fglm computation (not certified)
with AVX2 enabled, on an  Intel (R) Xeon (R) CPU
E7-4820v4 @ 2.00GHz, as reported by the msolve authors, they should be 
multipled by about 0.77 to account for the different frequencies. On
the other hand, CPU time for multi-threaded implementations are always
greater than for one-threaded implementations, especially for
relatively small computations or for computations requiring much
memory for each prime. For examples, Katsura 9 computation takes
3.06s CPU time with 1 thread instead of 6.16s with 16 threads (real time
1.33s), Katsura
10 takes 20s with 1 thread instead of 33s with 16 threads (real time 7.4s)
and Katsura 11 takes 170s with 1 thread instead of 245s with 16
threads (real time 46s).\\
For cp466, msolve authors report a running time of 71472s, but a value
of $d$ of 4096$\neq$728 that we obtain. Since our rur is certified,
there is no bug in giac, we suspect some mismatch in the data\\
\verb|gitlab.lip6.fr/eder/msolve-examples/-/raw/master/zero-dimensional/cp_d_4_n_6_p_6.ms|\\
The current
version of msolve does not support multiple CPUs, but it will
most certainly do in the near future.
\item The next 2 columns are CPU timings without certification and for
  certification.
\item the next column $Nd\sum \delta^2$ is the sum for all monomials
of the total degree squared, we will see that it gives a relatively good guess of the
certification execution time.
\item The RAM column is with certification (max certification threads
  6).
\item The $N$ column is the number of primes (these primes have a
  bitsize of 30), $\delta$ is the total degree of the initial system,
$d$ the dimension of the polynomials modulo the ideal, $l$ the number
of monomials of the system.
\item The last 2 columns give the time required to isolate all real
  roots of the minimial polynomial of the separating linear form.
The algorithm is a C++-transcription of Xcas user code sent by Alkis
Akritas (\cite{akritas2005comparative}). It is a little bit
parallelized, by running isolation of positive and negative real
roots in separate threads. It is most of the time at least one order of
magnitude faster than computing the polynomial, and is therefore
not a priority for further optimizations.
\end{itemize}
{\footnotesize
\begin{tabular}{|c|c|c|c||c||c|c|c|c||c|c|c|c||c|c|} \hline
        &  thrds  & giac &  giac & msolve & giac & cert &$Nd \cdot$ & & &&&&root&root\\
        & $t$ & real & cert & CPU & CPU & CPU & $\sum\delta^2$& RAM & $N$ & $\delta$ & $d$  &$l$& real  & CPU \\ \hline
Kat9* & 16 & 1.33 & 0.14 &8.41 & 6.16& 0.55& 5.2e6 & 67M & 84 & 2& 256& 74 &0.34 & 0.45\\
Kat10* & 16 & 7.4 & 0.53 & 43.7& 32.7 & 2.2 & 30e6 &225M & 194 & 2& 512& 90 &1.8 &  2.6 \\
Kat11* & 16 &  46 & 2.66 &424 & 245 &12& 0.15e9&781M &400 & 2& 1024& 107 &19.6 & 31.3 \\
Kat12* &16 & 550 & 18.1 &6262& 4726 & 87.8& 0.76e9 &3.5G & 862& 2& 2048& 126 &154 & 276 \\ 
Kat13* & 16 & 7320& 97& 0.89e5& 0.72e5 & 521 & 3.8e9&15.3G& 1831 &2 &4096 & 146 &1650 & 2900\\
Kat14* & 8 & 1.1e5 & 836 & 13e5& 7e5 & 3565 & 19e9&48G& 3980 &2& 8192& 168 &16e3 & 30e3 \\ \hline
Noon7 & 16 & 460 & 116 &4040& 4000 & 423& 1.2e9&3.6G& 1351 & 3 & 2173&64 &24.2 & 43.1 \\
Noon8 & 8 & 1.4e5 & 2160&6e5& 10e5& 8440& 14e9&33G& 4060 & 3&6545 & 81&413 & 814 \\
Phuoc & 16 & 363 & 5.3e5&4467& 4255 & 26e5& 6.1e12&37G & 781& 22&1102 & 32896&7.66 & 13  \\
Henr.6 & 16 & 6.08 & 10.5 &138& 54.3 & 35 & 0.15e9&492M & 310 & 6& 720& 69 &1.37 & 1.39 \\
Henr.7 &16 & 7200 & 3720&1.18e5& 0.87e5 & 0.12e5& 24e9& 33G & 2611& 7 & 5040& 134 &309  & 314 \\ \hline
Eco10 & 16 & 1.33 & 0.34&12.5& 10.2 & 1.22& 5.4e6 &108M & 57& 3 &256 & 64 &0.057 & 0.07 \\
Eco11 & 16 & 7.7 & 1.1 &90.3& 70.5 & 4.8 & 28e6 &287M& 119 & 3 & 512 & 76 &0.37& 0.39\\
Eco12 & 16 & 65 & 5.7 &877& 728 & 26.6 & 0.14e9&1.06G& 247 & 3 & 1024& 89 &2.52& 2.82\\
Eco13 & 16 & 715 & 35 &12137& 9340& 175 & 0.68e9 &9.5G& 509 & 3& 2048 & 103&25 & 27.5 \\
Eco14 & 16 & 0.1e5 & 186 &1.68e5 & 1.44e5 & 921 & 3.3e9&15.4G& 1048 & 3&4096 & 118&112 & 123  \\ \hline
cp352 & 16 & 2.7 & 26 &18.1& 18.5 & 148 & 0.95e9&253M & 338 & 4 & 288 & 866 &0.21 & 0.26 \\
cp362 & 16 & 38 & 694 & 311 & 429 & 3710& 21e9&1.5G & 1077 & 4 & 720 & 2265 & 1.25 & 1.75 \\
cp366 & 16 & 107 & 66& 255& 302 &380 & 2.1e9 &1.3G & 807 & 3& 729& 498&1.05 & 1.6 \\
cp372 & 16 & 1390 & 0.13e5& 9640& 0.15e5 & 0.69e5& 350e9&9.3G& 3143 & 4 & 1728 & 5187&52 & 91 \\
cp377 & 16 & 1125 & 3160& 0.12e5& 0.11e5 & 2030& 38e9&6.6G& 2892 & 3 & 2187 & 833&43 & 64 \\
cp382 & 8 & 0.32e5 & 2.1e5&2.7e5& 2.1e5 & 12e5& 4.8e12&51.3G & 8497& 4& 4032 & 10720&992 & 1655 \\
cp443 & 16 & 4.3 & 1821 &40.9& 33 & 1925 & 8.5e9&115M & 352 & 9 & 576& 922&0.53 & 0.64 \\
cp453 & 16 & 2.4e3& 3.3e5& 0.21e5& 0.27e5 & 13e5&4.7e12 &41G & 2843& 9&3456 & 8381&178 & 200 \\
cp466 & 16 & 240 & 342 &?& 930 & 1974& 8e9&2.6G & 3147 & 3 & 728 & 496 &5.11 & 7.01 \\
\hline 
\end{tabular}
}\\
Timings are given is seconds (sometimes rounded) with a relative precision of a few
percents (execution time depends on server load and RAM available). 
Large computation times are reported
with a \verb|1e5| exponent, this corresponds to a little more than 1
day (since one day=86400 seconds, i.e. \verb|0.864e5|).\\
This leads to the following observations~:
\begin{itemize}
\item 
If we plot the points ($x=$logarithm of $Nd\sum \delta^2$, $y=$logarithm of
  the certification execution time), we get 
\begin{center}
\includegraphics[width=0.8\textwidth]{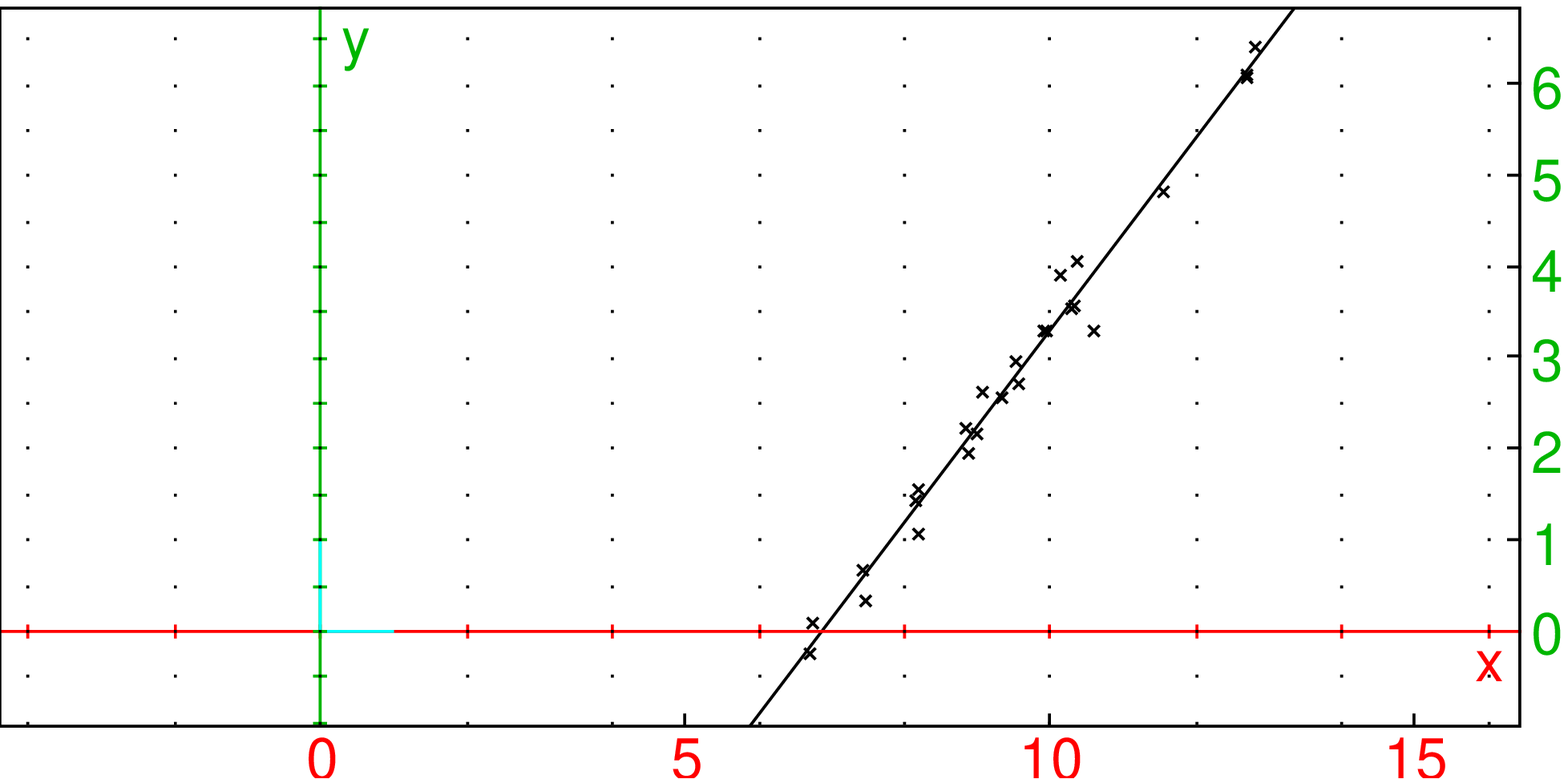}
\end{center}
The points are well grouped around the linear regression line (of
equation $y=1.06\cdot x-7.26$ with an $R^2=0.984$).
Therefore the
value $Nd\sum \delta^2$ (known after the probabilistic reconstruction
of the rur is done) gives a good guess of the order size of the
time that will effectively be required for certification.
\item One bad prime (over 1352) was observed for \verb|noon7|, $p=534856027$, and
  5 bad primes (over 4060) for \verb|noon8|.
\item Some examples above do not require as many primes as reported by
  msolve authors. I'm confident there is no
  bug inside Giac results since they are certified.
\item Except for noon7 and 8 where maple reported timings in \cite{berthomieu2021msolve}
  are better, the real multi-threaded timings of giac are currently
  the best available timings. Once msolve is multi-threaded, I expect that it should be a
  little bit faster than Giac for some examples.
\end{itemize}

Example of Giac/Xcas code:
\begin{verbatim}
threads:=16;
// debug_infolevel:=1;
rur_gbasis(1); // compute gbasis over Q (not the default)
// rur_certify(0); // do not certify rur 
kat10:=[x1 + 2*x2 + 2*x3 + 2*x4 + 2*x5 + 2*x6 + 2*x7 + 2*x8 + 2*x9
 + 2*x10 - 1, x1^2 + 2*x2^2 + 2*x3^2 + 2*x4^2 + 2*x5^2 + 2*x6^2 +
 2*x7^2 + 2*x8^2 + 2*x9^2 + 2*x10^2 - x1, 2*x1*x2 + 2*x2*x3 + 
 2*x3*x4 + 2*x4*x5 + 2*x5*x6 + 2*x6*x7 + 2*x7*x8 + 2*x8*x9 +
 2*x9*x10 - x2, x2^2 + 2*x1*x3 + 2*x2*x4 + 2*x3*x5 + 2*x4*x6 +
 2*x5*x7 + 2*x6*x8 + 2*x7*x9 + 2*x8*x10 - x3, 2*x2*x3 + 2*x1*x4 +
 2*x2*x5 + 2*x3*x6 + 2*x4*x7 + 2*x5*x8 + 2*x6*x9 + 2*x7*x10 - x4,
 x3^2 + 2*x2*x4 + 2*x1*x5 + 2*x2*x6 + 2*x3*x7 + 2*x4*x8 + 2*x5*x9 +
 2*x6*x10 - x5, 2*x3*x4 + 2*x2*x5 + 2*x1*x6 + 2*x2*x7 + 2*x3*x8 +
 2*x4*x9 + 2*x5*x10 - x6, x4^2 + 2*x3*x5 + 2*x2*x6 + 2*x1*x7 +
 2*x2*x8 + 2*x3*x9 + 2*x4*x10 - x7, 2*x4*x5 + 2*x3*x6 + 2*x2*x7 +
 2*x1*x8 + 2*x2*x9 + 2*x3*x10 - x8, x5^2 + 2*x4*x6 + 2*x3*x7 + 
 2*x2*x8 + 2*x1*x9 + 2*x2*x10 - x9];
vars:=[x1,x2,x3,x4,x5,x6,x7,x8,x9,x10];
time(H:=gbasis(kat10 ,vars,rur));
write("Hkat10",H); // use archive instead of write for fast read
// real root isolation
time(R:=realroot(eval(H[2],1)));
write("Rkat10",R);
size(R);
\end{verbatim}

\section{Conclusion}
We have now efficient probabilistic methods for rur computations
over $\Q$ and an efficient way to check it on the initial polynomial
system (except for dense systems of large degree), in other words
an efficient Las Vegas rur algorithm for radical ideals.
The question of an efficient deterministic algorithm is still open, it
may be impossible.

Some possible improvements are not implemented in Giac/Xcas~
\begin{itemize}
\item a more efficient (deterministic?) algorithm to find a linear
  separating form.
\item certifying with a recursive dense representation of the system
for dense polynomial equations.
\item certifying and a better implementation for non radical ideals.
\end{itemize}

\bibliography{gb.bib}

\begin{thebibliography}{10}

\bibitem{akritas2005comparative}
A.~G. Akritas and A.~W. Strzebonski.
\newblock A comparative study of two real root isolation methods.
\newblock {\em Nonlinear Analysis: Modelling and Control}, 10(4):297--304,
  2005.

\bibitem{Arnold2003403}
E.~A. Arnold.
\newblock {Modular algorithms for computing Gröbner bases }.
\newblock {\em Journal of Symbolic Computation}, 35(4):403 -- 419, 2003.

\bibitem{berthomieu2021msolve}
J.~Berthomieu, C.~Eder, and M.~{Safey El Din}.
\newblock {msolve: A Library for Solving Polynomial Systems}.
\newblock In {\em {2021 International Symposium on Symbolic and Algebraic
  Computation}}, 46th International Symposium on Symbolic and Algebraic
  Computation, Saint Petersburg, Russia, July 2021.

\bibitem{bouzidi2016solving}
Y.~Bouzidi, S.~Lazard, G.~Moroz, M.~Pouget, F.~Rouillier, and M.~Sagraloff.
\newblock Solving bivariate systems using rational univariate representations.
\newblock {\em Journal of Complexity}, 37:34--75, 2016.

\bibitem{chionh2002fast}
E.-W. Chionh, M.~Zhang, and R.~N. Goldman.
\newblock Fast computation of the bezout and dixon resultant matrices.
\newblock {\em Journal of Symbolic Computation}, 33(1):13--29, 2002.

\bibitem{faugere2017sparse}
J.-C. Faug{\`e}re and C.~Mou.
\newblock Sparse fglm algorithms.
\newblock {\em Journal of Symbolic Computation}, 80:538--569, 2017.

\bibitem{F99a}
J.-C. Faugère.
\newblock {A new efficient algorithm for computing Gröbner bases (F4).}
\newblock {\em Journal of Pure and Applied Algebra}, 139(1--3):61--88, June
  1999.

\bibitem{noro1999modular}
M.~Noro and K.~Yokoyama.
\newblock A modular method to compute the rational univariate representation of
  zero-dimensional ideals.
\newblock {\em Journal of Symbolic Computation}, 28(1-2):243--263, 1999.

\bibitem{parisse2013probabilistic}
B.~Parisse.
\newblock A probabilistic and deterministic modular algorithm for computing
  groebner basis over $\mathbb{Q}$.
\newblock {\em arXiv preprint arXiv:1309.4044}, 2013.

\bibitem{giac}
B.~Parisse and R.~{De Graeve}.
\newblock {Giac/Xcas computer algebra system, version 1.7.0-17}.
\newblock {\tt https://www-fourier.univ-grenoble-alpes.fr/$\tilde{\
  }$parisse/giac.html}, 2021.

\bibitem{rouillier1999solving}
F.~Rouillier.
\newblock Solving zero-dimensional systems through the rational univariate
  representation.
\newblock {\em Applicable Algebra in Engineering, Communication and Computing},
  9(5):433--461, 1999.

\bibitem{yap2000fundamental}
C.-K. Yap et~al.
\newblock {\em Fundamental problems of algorithmic algebra}, volume~49.
\newblock Oxford University Press Oxford, 2000.

\end{thebibliography}

\end{document}